\documentclass[aps,preprint,nofootinbib,floatfix,a4paper,prd]{revtex4-1}
\pdfoutput=1
\usepackage{graphicx}
\usepackage[latin1]{inputenc}
\usepackage{amsmath,amssymb}
\usepackage{hyperref}
\usepackage{epstopdf}
\usepackage{geometry}    

\def\be{\begin{equation}}
\def\ee{\end{equation}}
\def\bea{\begin{eqnarray}}
\def\eea{\end{eqnarray}}

\begin{document}

\title{Electron Electric Dipole Moment\\
in Mirror Fermion Model\\
with Electroweak Scale Non-sterile\\
Right-handed Neutrinos}

\author{Chia-Feng Chang}
\email[]{a29788685@gmail.com}
\affiliation{Department of Physics, National Taiwan University, Taipei 116, Taiwan}

\author{P. Q. Hung}
\email[]{pqh@virginia.edu}
\affiliation{Department of Physics, University of Virginia, Charlottesville, VA 22904-4714, USA\\
and\\
Center for Theoretical and Computational Physics, Hue University College of Education, Hue, Vietnam}

\author{Chrisna Setyo Nugroho}
\email[]{setyo13nugros@gmail.com}
\affiliation{Department of Physics, National Taiwan Normal University, Taipei 116, Taiwan}

\author{Van Que Tran}
\email[]{apc.tranque@gmail.com}
\affiliation{Department of Physics, National Taiwan Normal University, Taipei 116, Taiwan}

\author{Tzu-Chiang Yuan}
\email[]{tcyuan@phys.sinica.edu.tw}
\affiliation{Institute of Physics,\\ Academia Sinica,\\ Nangang, Taipei 11529, Taiwan\\
and\\
Physics Division,\\ National Center for Theoretical Sciences, Hsinchu, Taiwan}

\date{\today}                                          

\begin{abstract}
The electric dipole moment of the electron is studied in detail in an extended mirror fermion model 
with the following unique features of (a) right-handed neutrinos are non-sterile and have masses 
at the electroweak scale, and (b) a horizontal symmetry of the tetrahedral group is used
in the lepton and scalar sectors. 
We study the constraint on the parameter space of the model imposed by the latest ACME experimental limit on
electron electric dipole moment.
Other low energy experimental observables such as
the anomalous magnetic dipole moment of the muon,
charged lepton flavor violating processes like muon decays into electron plus photon and muon-to-electron
conversion in titanium, gold and lead are also considered in our analysis for comparison.
In addition to  the well-known CP violating Dirac and Majorana phases in the neutrino mixing matrix, 
the dependence of additional phases of the new Yukawa couplings in the model
is studied in detail for all these low energy observables.

\end{abstract}

\maketitle

\section{Introduction}
\label{sec1}
 
The particle spectrum of the Standard Model (SM) has now been completed by the discovery of the 125 GeV Higgs
boson at the Large Hadron Collider  (LHC). Nevertheless, many questions remained unanswered 
within the SM. On the conceptual side, we do not understand the instability of the Higgs boson mass under quantum corrections, indicating that SM is very sensitive to new physics beyond the TeV scale; 
while on the phenomenological side, we have issues like 
the Baryon Asymmetry of the Universe (BAU), dark matter, and neutrino masses etc.
The current popular view is that SM is just a low energy effective theory of a better one at a higher scale with new physics that can address some or all of the above issues. Indeed many beautiful ideas had been suggested in the literature to solve some of these issues.
Current LHC constraint is already quite stringent on the scale of new physics 
$\Lambda_{\rm NP} \sim$  a half to a few TeV, should the new physics 
be supersymmetry or extra dimension or sequential fourth generation, or technicolor etc.
While one should continue the direct searches for new particles at the LHC, 
looking for new physics indirectly from low energy observables where 
new particles only exist virtually at the loop level is an important alternative avenue.
Historically one can recall that the charm quark was predicted long before its discovery
by the GIM mechanism~\cite{Glashow:1970gm}, which was engaged to suppress flavor changing neutral currents in the box diagrams of the $K \overline K$ kaon system.

The electric dipole moment (EDM) of an elementary particle
is one such low energy observable which is sensitive to 
new CP violating phases from new physics. 
As is well known the CP violation phase in the 
Cabibbo-Kobayashi-Maskawa (CKM) quark mixing matrix is too minuscule to account for the 
BAU, characterized by the ratio of the net baryon number density to the entropy density 
in the Universe~\cite{planck2015}, 
$$
Y_B \equiv n_B / s = (8.61 \pm 0.09 ) \times 10^{-11} \; .
$$ 
Moreover, the SM contribution to the electron EDM from the CKM CP violation phase must
arise at least at the four-loop level~\cite{Shabalin:1978rs,Shabalin:1979mw,EDM:SM,Booth:1993af}. 
The reasons are as follows: 
Due to the structure of the particle exchange symmetry in the loop integrals of 
the various diagrams, the $W$ boson EDM vanishes 
to two-loop order in SM model, but it can be non-vanishing with one more gluon-dressed loop. 
By attaching the two external $W$ boson lines of the three-loop diagrams 
to the electron one can generate the electron EDM in SM.
Thus the resulting electron EDM ($d_e$) in SM is a four-loop result, 
estimated to be  $\sim 8 \times 10^{-41} e \cdot {\rm cm}$~\cite{Booth:1993af}, which is twelve orders of magnitude
below the current experimental limit (see below). 
Therefore a positive measurement of the electron EDM
at the current sensitivities of various experiments or their projected improvements in the near future 
would definitely imply new sources of CP violation. 
New CP violating phases might then be helpful to solve the BAU puzzle.

The latest measurement of the electron EDM was done by the ACME Collaboration~\cite{Baron:2013eja}
using the polar molecule thorium monoxide (ThO) just a few years back,
\be
d_e = (-2.1 \pm 3.7_{\rm stat} \pm 2.5_{\rm syst} ) \times 10^{-29} \; e \cdot {\rm cm} \; .
\ee
This corresponds to a 90\% confidence limit,
\be
\vert d_e \vert < 8.7 \times 10^{-29} \; e \cdot {\rm cm} \; ,
\ee
which is an improvement by a factor of 12 over the previous best measurements.

In this work, we study the electron EDM in a class of mirror fermion model proposed some time ago 
by one of the authors \cite{Hung:2006ap}. 
We will demonstrate that the above ACME limit can put stringent constraints on the parameter space of the
extended mirror fermion model discussed below.

Here we briefly review the salient features of the original mirror model~\cite{Hung:2006ap}.
In contrast with the left-right symmetric models, the gauge group was chosen to be the same as SM, 
only mirror fermions were introduced. 
Right-handed neutrinos were introduced as well,
but instead of being sterile singlets, they were put inside right-handed weak doublets 
with mirror charged leptons for each generation. In addition to the SM Higgs doublet, 
the Georgi-Machacek (GM) triplets \cite{Georgi:1985nv,Chanowitz:1985ug} were also needed to provide Majorana masses 
for right-handed neutrinos. To obtain the correct electroweak symmetry breaking pattern,
the triplet vacuum expectation value (VEV) should be around the electroweak scale as well.
Thus the non-sterile right-handed neutrinos have Majorana masses of the order of
electroweak scale which imply immediate consequences at the LHC! 
Furthermore, an electroweak scalar singlet was also bought into the model to generate 
tiny Dirac neutrino masses of order eV through small enough VEV and 
provide very small mixings between SM fermions and their mirrors.

Recently, many phenomenological implications of the mirror model~\cite{Hung:2006ap} have been explored further.
We summarize what we have been done in a series of works involving 
various collaborations:
In~\cite{Hoang:2013jfa}, the model was challenged by the electroweak precision measurements. 
It was shown that the dangerously large contributions to the oblique parameters from the mirror fermions 
(especially the $S$ parameter)
can be tamed by the opposite contributions from the Higgs triplets. 
In~\cite{Hoang:2014pda}, the original mirror model was extended by adding a mirror Higgs doublet so as 
to accommodate the LHC data for the SM Higgs signal strengths of various channels.
Searches for mirror fermions at the LHC were studied in~\cite{Chakdar:2015sra} for mirror quarks
and~\cite{Chakdar:2016adj} for mirror leptons.
In~\cite{Hung:2015nva}, 
the neutrino and charged lepton masses and mixings were discussed in the mirror model with 
a horizontal $A_4$ symmetry imposed on the lepton sector.
Subsequently, in~\cite{Hung:2015hra}, the charged lepton flavor violating (CLFV) 
radiative decay $\mu \to e \gamma$ was studied in details  
in this mirror model with the $A_4$ symmetry extension, 
updating an earlier calculation~\cite{Hung:2007ez} done for the original model.
Moreover, the $\mu -e$ conversion in nuclei was also studied~\cite{Hung:2017voe}.
In~\cite{Chang:2016ave}, the CLFV Higgs decay $h (125 \, {\rm GeV}) \to \mu \tau$ was studied 
for the extended mirror model with a mirror Higgs doublet~\cite{Hoang:2014pda} .

There exists a huge amount of studies of the electron EDM in the literature for 
other new physics models with CP violation, for example, the generic two Higgs doublet model \cite{Jung:2013hka}, 
minimal supersymmetric standard model (MSSM)~\cite{Nath:1991dn,Ibrahim:1997gj,Ellis:2008zy,Ibrahim:2014oia}, 
models with sterile neutrinos~\cite{Abada:2015trh}, {\it etc.}
For recent reviews on this topics, see for example \cite{Pospelov:2005pr,Engel:2013lsa,Chupp:2014gka,LinderPlastscherQueiroz}.
We focus on the electron EDM in the mirror model with the $A_4$ symmetry 
as discussed in~\cite{Hung:2015nva}.
      
This paper is organized as follows. 
In Sect.~\ref{sec2}, we give some more details of the model 
by spelling out the relevant lepton and scalar spectra, their $A_4$ assignments, and
the new Yukawa couplings.
In Sect.~\ref{sec3}, we present the formulas of the lepton EDMs.
In Sect.~\ref{sec4}, we first discuss the assumptions and scenarios used in the numerical analysis and
then present the numerical results for the electron EDM.
We conclude in Sect.~\ref{sec5}. 
Some useful formulas are relegated to the end by an Appendix. 

\section{Brief Review of the Model}
\label{sec2}
 
In this section, we highlight the original mirror fermion model discussed in~\cite{Hung:2006ap} and its recent $A_4$ extension \cite{Hung:2015nva}.

For each generation $i$, the SM lepton doublet $l_{Li}$ and singlet $e_{Ri}$ are accompanied with 
mirror fields $l^M_{Ri}$ and $e^M_{Li}$ respectively.
For the scalars, $\Phi_M$ is the mirror Higgs doublet of $\Phi$, both have hypercharge
$Y/2 = 1/2$; $\xi$ and $\tilde \chi$ are the 
two GM triplets with $Y/2$ equal 0 and 1 respectively;
and $\phi_{0S}$ and $\phi_{iS}$ $(i=1,2,3)$ are all singlets.

Recall that the tetrahedron symmetry group $A_4$ has four irreducible representations $\bf 1$, $\bf 1'$, $\bf 1''$, 
and $\bf 3$ with the following multiplication rule
${\bf 3} \times {\bf 3}  =    {\bf 3_1} (23, 31, 12) + {\bf 3_2} (32, 13, 21) 
+ {\bf 1} (11+ 22 + 33) + {\bf 1'} (11 + \omega^2 22 + \omega 33) 
+ {\bf 1''} (11 + \omega 22 + \omega^2 33)$
where $\omega = e^{2 \pi i/3} = -\frac{1}{2} + i \frac{\sqrt 3}{2}$.
Note $\omega^2 = \omega^*$.
For the $A_4$ assignments,  
$l^M_{Ri}$, $e^M_{Li}$, $l_{Li}$, $e_{Ri}$ and $\phi_{iS}$ are triplets while 
all other fields are singlets.

In the gauge eigenbasis (fields with superscript 0), one can write down the following $A_4$ 
invariant Yukawa couplings,
\bea
- {\cal L}_S 
& = & g_{0S} \phi_{0S} (\overline{l^0_L}   l^{0M}_R)_{\bf 1} 
+ g_{1S} \vec \phi_S \cdot (\overline{l^0_L} \times l_R^{0M})_{\bf 3_1}
+ g_{2S} \vec \phi_S \cdot (\overline{l^0_L} \times l_R^{0M})_{\bf 3_2} 
\nonumber \\
&+&  g^\prime_{0S} \phi_{0S} (\overline{e^0_R}   e^{0M}_L)_{\bf 1} 
+ g^\prime_{1S} \vec \phi_S \cdot (\overline{e^0_R} \times e_L^{0M})_{\bf 3_1}
+ g^\prime_{2S} \vec \phi_S \cdot (\overline{e^0_R} \times e_L^{0M})_{\bf 3_2} 
+ {\rm H.c.} \;\;
\label{A4Lang1}
\eea
The singlet scalars $\phi_{0S},\vec\phi_{S}=(\phi_{1S},\phi_{2S},\phi_{3S})$ are the only fields connecting the SM fermions and their mirror counterparts.
After the scalar singlets develop small VEVs $v_0 =\langle \phi_{0S} \rangle$ 
and $v_i =\langle \phi_{iS} \rangle$ of order $10^{5}$ eV, one obtains the tiny neutrino Dirac mass~\cite{Hung:2006ap}.

On the other hand, the Majorana mass term for the non-sterile right-handed neutrinos can be generated by
the following $A_4$ invariant Lagrangian~\cite{Hung:2006ap,Hung:2015nva}
\be
{\mathcal L}_M = g_M \left( l^{M,T}_{R} \sigma_2 \right)
\left( i \tau_2 \tilde \chi \right) l^M_{R} + {\rm H.c.} \; .
\label{LagMajorana}
\ee
When the neutral component of the $A_4$ singlet $\tilde \chi$ develops a VEV 
$\langle \chi_0 \rangle = v_M\sim v_{\rm SM}= 246$ GeV, one obtains
the Majorana mass at the electroweak scale~\cite{Hung:2006ap}.

In terms of the physical mass eigenstate fields, the Yukawa couplings in (\ref{A4Lang1}) read
\bea
\label{La}
{\mathcal L}^{l}_{S} & = & - \sum_{k=0}^3 \sum_{i,m=1}^3  \left( \bar{l}_{Li} \, {\cal U}^{L \, k}_{im} l^M_{R m}  
+ \bar{ e}_{R i}  \, {\cal U}^{R \, k}_{im} e^M_{Lm} \right) \phi_{kS} + {\rm H.c.} \; ,
\eea
where we have grouped the singlet and triplet scalars $\phi_{0S}$ and $\vec \phi_S$ into 
$\phi_{kS}$ with $k=0,1,2,3$.
The coupling coefficients ${\cal U}^{L \, k}_{im}$ and ${\cal U}^{R \, k}_{im}$ are given by~\cite{Hung:2015hra}
\bea
\label{UL}
{\cal U}^{L \, k}_{im} 
& \equiv & \left ( U^\dagger_{\rm PMNS}  \cdot   M^k \cdot  U^{M}_{\rm PMNS} \right)_{im} \;\; , \nonumber \\
& = & \sum_{j,n = 1}^3 \left( U^\dagger_{\rm PMNS} \right)_{i j}   M^k_{jn}  
\left(  U^{M}_{\rm PMNS} \right)_{nm} \; \; , \\
{\cal U}^{R \, k}_{im} 
& \equiv & \left( U^{\prime \, \dagger}_{\rm PMNS} \cdot   M^{\prime \, k} \cdot U^{\prime \, M}_{\rm PMNS} \right)_{im} \; \; , \nonumber \\
& = & \sum_{j,n = 1}^3 \left( U^{\prime \, \dagger}_{\rm PMNS} \right)_{i j}   M^{\prime \, k}_{jn}  
\left(  U^{\prime \, M}_{\rm PMNS} \right)_{nm} \; \; ,
\label{UR}
\eea
where the matrix elements for the auxiliary matrices $M^k_{jn}$ 
and $M^{\prime \, k}_{jn}$ with $k=0,1,2,3$ depend on the new Yukawa couplings.
Their definitions can be found in~\cite{Hung:2015hra}.
$U_{\rm PMNS}$ is the usual neutrino mixing matrix defined as
$U_{\rm PMNS}=  U_{\nu}^{\dagger} U^{l}_{L} = U_{\rm CW} U^{l}_{L}$
with~\cite{Hung:2015hra}
\be
\label{UCW}
U_\nu =  U^\dagger_{\rm CW} =
\frac{1}{\sqrt{3}}
\left(
  \begin{array}{cccc}
  1 & 1 & 1 \\
  1 & \omega^2 & \omega \\
  1 & \omega & \omega^2 \\
  \end{array}
\right) \; ,
\ee
and $U^l_L$ is the unitary matrix that diagonalizes the charged lepton mass matrix.
In the $A_4$ limit, $U^l_L = 1$. In~\cite{Hung:2015nva}, it was parameterized as a Wolfenstein-like
matrix with elements whose values are constrained by the experimental input of $U_{\rm PMNS}$.
Analogously the mirror and right-handed counter-parts  of $U_{\rm PMNS}$ are 
$U^{M}_{\rm PMNS} $, $U^\prime_{\rm PMNS}$ 
and $U^{\prime M}_{\rm PMNS} $ defined  as
$U^{M}_{\rm PMNS} = U^{\dagger}_{\nu} U^{l^M}_R$, 
$U^\prime_{\rm PMNS}=  U_{\nu}^{\dagger} U^{l}_{R}$,
and
$U^{\prime M}_{\rm PMNS} = U^{\dagger}_{\nu} U^{l^M}_L$.
Certainly, among these four PMNS-type mixing matrices, only $U_{\rm PMNS}$ has been determined 
experimentally.

\begin{figure}[hptb!]
\centering
\includegraphics[width=0.85\textwidth]{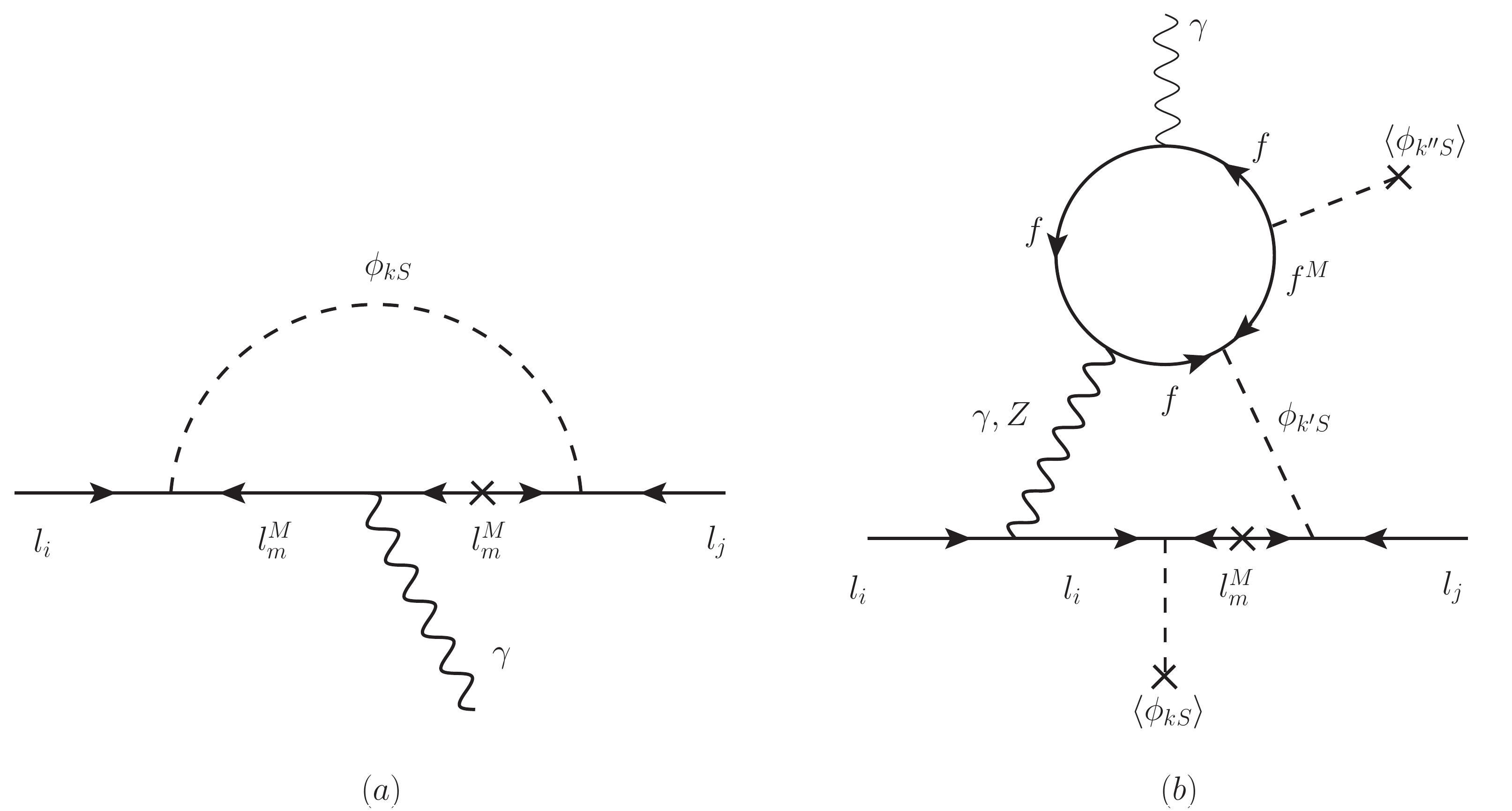} 
\caption{\small Feynman diagrams contributing to charged lepton EDM in mirror fermion model. 
(a) one-loop diagram and (b) two-loop Zee-Barr type diagram.} 
\label{eEDMdiagram}
\end{figure}

 \section{Charged Lepton Electric Dipole Moments} 
 \label{sec3}

The electric dipole moment (EDM) operator is defined as
\begin{equation}
\label{LeEDM}
 {\cal L}_{l_i}^{\rm EDM}=-i\frac{d_{l_{i}}}{2} \,\bar{l}_{i}\,{\sigma}^{\mu\nu}\,{\gamma}_5\,{l}_i\,F_{\mu\nu}\,,
 \end{equation}
 where  $F_{\mu\nu}$ is the electromagnetic field strength and the coefficient $d_{l_{i}}$ is the electric dipole moment
 for the $i$-th generation charged lepton $l_i$.

The one-loop and two-loop Feynman diagrams contributing to the charged lepton EDMs 
in the mirror model that we are discussing are depicted in Fig.~\ref{eEDMdiagram}. 
The two-loop Zee-Barr type diagram~\cite{Barr:1990vd} in Fig.~\ref{eEDMdiagram}b is completely negligible 
due to the mixings between SM fermions and their mirrors
which are proportional to the very small VEVs of the singlets.
Moreover it is suppressed by two more powers of the small new Yukawa couplings as compared with the one-loop diagram in Fig.~\ref{eEDMdiagram}a.
We will focus on the one-loop diagram in Fig.~\ref{eEDMdiagram}a. The one-loop amplitude for the 
process $l^-_i (p) \to l^-_j (p') + \gamma (q) $ has been computed in \cite{Hung:2015hra} 
with the following matrix element,
\be
\label{M1l}
{\cal M}\left( l^-_i \to l^-_j \gamma \right) = \epsilon^*_\mu(q) \bar u_j (p')  \left\{ 
i \sigma^{\mu \nu} q_\nu \left[ C^{ij}_L P_L + C^{ij}_R P_R \right] \right\} u_i(p) \; \; ,
\ee
where $P_{L,R} = (1 \mp \gamma_5 )/2$ are the chiral projection operators and the coefficients $C^{ij}_{L,R}$
are summarized in the Appendix for convenience.

The amplitude $\cal M$ in Eq.~(\ref{M1l}) can be reproduced by the following effective interaction
\begin{equation}
{\mathcal L}_{\rm eff} = - \,\frac{1}{2} \,\bar l_j  \left\{ 
i \sigma^{\mu \nu}\left[ C^{ij}_L P_L + C^{ij}_R P_R \right] \right\} l_i \, F_{\mu\nu}\; \; .
\end{equation}
Comparing with the lepton  EDM Lagrangian in Eq.~(\ref{LeEDM}), we can extract the electric dipole moment 
$d_{l_{i}}$ as~\cite{Hung:2015hra}
\bea
d_{l_i} & = & \frac{i}{2} \left( C^{ii}_L - C^{ii}_R \right) \;\; , \nonumber \\
&= & + \frac{e}{16 \pi^2}  
\sum_{k=0}^3 \sum_{m=1}^3 \frac{1}{m_{{l^M_m}}}  \mathrm{Im} \left( \mathcal{U}^{L\, k}_{im} 
\left( \mathcal{U}^{R\, k}_{im} \right)^* \right)
{\cal J} \left( \frac{m^2_{\phi_{kS}}}{m^2_{{l^M_m}}} \right)   \;\; ,
\label{edm2}
\eea
where $m_{l^M_m}$ and $m_{\phi_{kS}}$ are mirror leptons and scalar Higgs masses respectively, 
${\mathcal U}^{L \, k}_{im}$ and ${\mathcal U}^{R \, k}_{im}$ 
are defined in (\ref{UL}) and (\ref{UR}), and ${\mathcal J}(r)$ is a loop function defined in the Appendix (Eq.~(\ref{J})).
We note that the dependence of new Yukawa couplings are hidden in the auxiliary matrices 
$M^k_{jn}$ and $M^{\prime \, k}_{jn}$ in the definitions of ${\mathcal U}^{L \, k}_{im}$ and ${\mathcal U}^{R \, k}_{im}$ 
given in (\ref{UL}) and (\ref{UR}) respectively.

We note the contribution from the mirror fermion-scalar singlet loop to the charged lepton EDM is similar to the 
chargino-sneutrino loop in MSSM~\cite{Ibrahim:2014oia}. One can check that 
our EDM expression in (\ref{edm2}) is consistent with the MSSM result given by (11) in~\cite{Ibrahim:2014oia}.

\section{Analysis}
\label{sec4}

We will first discuss the parameter space that is relevant in our numerical analysis. Our approach here is similar to those 
adopted in previous works~\cite{Hung:2015hra,Chang:2016ave}.

\begin{itemize}

\item
The six new Yukawa couplings $g_{0S}$, $g_{1S}$, $g_{2S}$, $g'_{0S}$, $g'_{1S}$, $g'_{2S}$ are in general complex. 
Recall that we have the following relations $g_{2S} = ( g_{1S} )^*$ and $g^\prime_{2S} = (g^{\prime}_{1S})^*$
due to the reality of the eigenvalues of the Dirac neutrino mass matrix~\cite{Hung:2015nva}.
Thus we will write the couplings as follows: 
 \be
 g_{0S} = |g_{0S}|\,e^{i\delta_{0}}, g_{1S} = |g_{1S}|\,e^{i\delta_{1}} , g^{\prime}_{0S} = |g^{\prime}_{0S}|\,e^{i\delta^{\prime}_{0}}, g^{\prime}_{1S} = |g^{\prime}_{1S}|\,e^{i\delta^{\prime}_{1}} \; .
 \ee
The new phases in these Yukawa couplings are the new sources of CP violation. 
Note that one can absorb the phase $\delta_0$ of $g_{0S}$ in the first term in (\ref{A4Lang1}) by redefinition of $\phi_{0S}$.
However, it will show up at the fourth term as $e^{i (\delta^\prime_0 - \delta_0)}$. 
Similarly, one can absorb the phase $\delta_{1}$ of $g_{1S}$  in the second term in (\ref{A4Lang1})
by redefinition of $\vec \phi_{S}$ but it will reappear at the fifth 
term as $e^{i (\delta'_1 - \delta_1)}$. 
Since $g_{2S} = ( g_{1S} )^*$ and $g^\prime_{2S} = (g^{\prime}_{1S})^*$, similar comment can be made for the
third and the sixth terms in (\ref{A4Lang1}).
Thus one expects any physical observable must be depending 
on the phase differences 
\be
\label{alphaandbeta}
\alpha = \delta_0 - \delta'_0 \;\;\; , \;\;\; \beta = \delta_1 - \delta'_1 \; .
\ee
For $\delta_{0,1}$ and $\delta^\prime_{0,1}$ range from 0 to $2 \pi$, $\alpha$ and $\beta$ range from 
$-2 \pi$ to $2 \pi$.
 
\item 

For the PMNS matrix it is most commonly parameterized as~\cite{PDG-neutrinos}
\bea
U_{\rm PMNS} = \left( 
\begin{array}{ccc}
c_{12}c_{13} & s_{12}c_{13} & s_{13} e^{-i\delta_{\rm CP}} \\
-s_{12}c_{23}-c_{12}s_{23}s_{13}e^{i \delta_{\rm CP}} & c_{12}c_{23}-s_{12}s_{23}s_{13}e^{i \delta_{\rm CP}} & s_{23}c_{13}\\
s_{12}s_{23}-c_{12}c_{23}s_{13}e^{i \delta_{\rm CP}} & -c_{12}s_{23}-s_{12}c_{23}s_{13}e^{i \delta_{\rm CP}} & c_{23}c_{13}
\end{array} \right) \cdot P  \nonumber 
\eea
where $s_{ij} \equiv \sin \theta_{ij}$,  $c_{ij} \equiv \cos \theta_{ij}$ with $\theta_{ij} \in [0,\pi/2]$ being 
the mixing angles, $\delta_{\rm CP} \in [0, 2 \pi]$ being the CP-violating Dirac phase, and 
$P = {\rm diag} [ 1 , e^{i \alpha_{21}/2} , e^{i \alpha_{31}/2} ]$ is the Majorana phase matrix. 
While the current neutrino experimental data are not sensitive to these Majorana phases,
one can show that the electron EDM is independent of these phases in the
mirror fermion model that we are studying.
We will set $\delta_{\rm CP} = - \pi/2$ (or equivalently $3 \pi /2$) 
as suggested by recent data of the different appearance rates 
for $\nu_\mu \to \nu_e$~\cite{IwamotoICHEPTalk,Adamson:2016tbq} and $\bar \nu_\mu \to \bar \nu_e$~\cite{IwamotoICHEPTalk}, 
as well as $\bar \nu_\mu$ disappearance rate~\cite{Adamson:2016xxw} in various experiments,
which is consistent with the most recent global analysis of neutrino 
oscillation data~\cite{Gonzalez-Garcia:2015qrr, Esteban:2016qun}.
The current global fit results of three mixing angles were given in Table 1 
of Ref.~\cite{Esteban:2016qun}.
For convenience, we list them in Table~\ref{mixingparameters} here.

\begin{table}[hptb!]
\caption{The current global fit results ($\pm 1\sigma$)
of three mixing angles taken from~\cite{Esteban:2016qun}}
\begin{tabular}{|c||c|c|}
\hline
Mixing angles  & Normal Hierarchy (best fit) & Inverted Hierarchy ($\Delta\chi^2 = 0.83$)\\
\hline
\hline
$\sin^2\theta_{12}$ & $0.306^{+0.012}_{-0.012} $ & $0.306^{+0.012}_{-0.012} $\\
$\sin^2\theta_{23}$ & $0.441^{+0.027}_{-0.021}$ & $0.587^{+0.020}_{-0.024}$\\
$\sin^2\theta_{13}$ & $0.02166^{+0.00075}_{-0.00075}$ & $0.02179^{+0.00076}_{-0.00076}$\\
\hline
\end{tabular}
\label{mixingparameters}
\end{table}

\item For the three unknown PMNS matrices we assume that they are equal to each other 
and study the following two scenarios: 
\begin{itemize}
\item{Scenario A}
$$
U^{M}_{\rm PMNS} = U^\prime_{\rm PMNS} =U^{\prime M}_{\rm PMNS} = U_{\rm CW}
$$

\item{Scenario B}
$$U^{M}_{\rm PMNS} = U^\prime_{\rm PMNS} =U^{\prime M}_{\rm PMNS} =U_{\rm PMNS}$$
\end{itemize}

\item
For the masses of the singlet scalars $\phi_{kS}$, we assume 
\be
\label{mS}
m_{\phi_{0S}} \approx m_{\phi_{kS}} = m_S = 1 \; {\rm GeV} \; .
\ee
Recall that in the seesaw mechanism the light neutrino mass is 
$m_\nu^{\rm light} \sim (m_\nu^D)^2 / M_R$. If the neutrino Dirac mass $m_\nu^D$
is generated at the electroweak scale, one must require the right-handed neutrino mass 
scale $M_R$ to be at the grand unification scale
in order to achieve light neutrino mass $m_\nu^{\rm light}$ of order eV. 
However, in the electroweak scale seesaw mechanism~\cite{Hung:2006ap}, 
$m^D_\nu \sim g_S \langle \phi_S \rangle$ and 
$M_R$ is of the order electroweak scale.
To achieve eV light neutrino mass, one needs 
$$g_S \langle \phi_S \rangle \sim \sqrt{M_R/{\rm TeV}} \times  {\rm MeV}.$$
Thus for $M_R \sim v_{\rm SM} = 246$ GeV, as $g_S$ varies from 
$5 \times 10^{-4}$ to $1$, 
$\langle \phi_S \rangle$ varies from 1 GeV to 0.5 MeV.
The mass of the singlet $m_{\phi_S} \sim \lambda_S \langle \phi_S \rangle$ where $\lambda_S$ is a generic quartic coupling of order one in the scalar potential. 
So we choose the common mass $m_S$ to be 1 GeV in (\ref{mS}) as a nominal value.

Similarly, for the mirror lepton masses, we assume they are degenerate, {\it i.e.} 
\be
m_{l^M_k}  = m_{M} \; ,
\ee
and vary the common mass $m_M$ from 100 to 800 GeV. Thus $m_M \gg m_S$, the loop function 
${\cal J} ( m^2_{\phi_{kS}} / m^2_{l^M_m} ) \approx {\cal J}(0) = 1/2$ which is not sensitive to the masses of 
the singlets and the mirror leptons. 

With these assumptions, the electric dipole moment in (\ref{edm2}) can be simplified as
\be
d_{l_i} \approx + \frac{e}{32 \pi^2} \frac{1}{m_M} J_{l_i}\; ,
\label{edmsimple}
\ee
where
\be
 J_{l_i} \equiv \mathrm{Im} 
\sum_{k=0}^3 \sum_{m=1}^3  \mathcal{U}^{L\, k}_{im} \left( \mathcal{U}^{R\, k}_{im} \right)^* \; .
\label{Ji}
\ee
One can easily show that $J_e$ is independent of the Majorana phases while $J_\mu$ and $J_\tau$ in general do.

For Scenario A, $J_e$ can be expressed as
\be
J^A_e = |g_{0S}|\,|g^{\prime}_{0S}| \, \Big( C_{1}\sin(\alpha)+ C_{2}\sin (\delta_{\rm CP}-\alpha)\Big) 
 + \,2 \, |g_{1S}|\,|g^{\prime}_{1S}|\,C_{2}\sin(\delta_{\rm CP})\cos(\beta) \,
\label{JA}
\ee
with
\bea
\label{alp}
C_{1}&=& \frac{1}{\sqrt{3}}\,\big[c_{12} c_{13}+s_{12}(s_{23}-c_{23})\big]\,,\nonumber \\
C_{2}&=& \frac{1}{\sqrt{3}}\,c_{12} s_{13}(s_{23}+c_{23})\,.
\eea
$J_e^A$ achieves its extremum at 
$\tan \alpha = -\cot \delta_{\rm CP} + \frac{C_1}{\sin \delta_{\rm CP} \, C_2}$ and $\sin \beta=0$. Numerically, $|J_e^A|$  
and therefore $|d_e|$ is maximized at $\alpha \approx 4.93$ ($4.95$) for normal (inverted) hierarchy and $\beta =0$.
For Scenario B, we simply have 
\bea
J^B_e =  |g_{0S}|\,|g^{\prime}_{0S}|\,\sin(\alpha)\,.
\label{JB}
\eea
Note that in Scenario B, only the magnitudes of the $A_4$ singlet couplings $g_{0S}$ and $g'_{0S}$ and their relative phase $\alpha$ entered in (\ref{JB}), the Dirac phase $\delta_{\rm CP}$ and the $A_4$-triplet couplings $g_{1S}$ and $g'_{1S}$ do not contribute! Clearly, $J_e^B$ is maximized at $\alpha=\pi/2$ such that $d_e$ has its largest value.

\item 
For comparisons, we include in our analysis the CLFV processes $\mu-e$ conversion in nuclei (titanium, gold and lead) and $\mu \to e \gamma$, and 
the muon anomalous magnetic dipole moment (MDM).

The present experimental upper limits on the branching ratios of $\mu-e$ conversion for  nuclei  titanium \cite{Dohmen:1993plB}, gold \cite{Bertl:2006up} and lead~\cite{Honecker:1996zf} are 
\bea
  {\rm Br}(\mu^- + {\rm Ti}  \to e^- + {\rm Ti}) & < & 4.3 \times 10^{-12} \; (90 \% \, {\rm C.L.)} \,{\rm [SINDRUM \, II]} 
  \;, \\
  {\rm Br}(\mu^- + {\rm Au} \to e^- + {\rm Au}) & < & 7 \times 10^{-13} \; (90\% \, {\rm C.L.}) \,{\rm [SINDRUM \, II]} \; , \\
  {\rm Br}(\mu^- + {\rm Pb} \to e^- + {\rm Pb}) & < & 4.6 \times 10^{-11} \; (90\% \, {\rm C.L.}) \,{\rm [SINDRUM \, II]} \; .
  \eea
And the projected sensitivities for aluminum and titanium are
~\cite{Bartoszek:2014mya, COMETphase1, Kuno:2013mha, Knoepfel:2013ouy, Barlow:2011zza}
\bea
\label{COMETmu2eprojected}
  {\rm Br}(\mu^- + {\rm Al}  \to e^- + {\rm Al}) & < & 3 \times 10^{-17} \,\hspace{10pt}{\rm(Mu2e, COMET)}\, ,\\
  \label{mu2eIIPRISMprojected}
  {\rm Br}(\mu^- + {\rm Ti}  \to e^- + {\rm Ti}) & < &  10^{-18} \,\hspace{10pt}{\rm(Mu2e \, II, PRISM)}\, .
\eea

The current limit \cite{TheMEG:2016wtm} and projected sensitivity \cite{Renga:2014xra} for 
${\rm Br}(\mu \rightarrow e \gamma)$  from the MEG experiment are
   \bea
      {\rm Br}(\mu \to e \gamma) & \le & 4.2 \times 10^{-13} \; (90\% \, {\rm C.L.}) \,\hspace{10pt} {\rm[MEG, \, 2016]}  \, ,\\
      {\rm Br}(\mu \to e \gamma) & \sim & 4 \times 10^{-14} \;\hspace{30pt}{\rm[Projected \; Sensitivity]} \, .
   \eea

For the muon anomalous magnetic dipole moment, we have from the E821 experiment~\cite{muonanomaly}
the $3.6 \sigma$ discrepancy between the measurement and the SM prediction
\be
\label{E821}
\Delta a_\mu  \equiv   a^{\rm exp}_{\mu} - a^{\rm SM}_{\mu}   =  288(63)(49) \times 10^{-11} \; , 
\ee
where the first errors are experimental and the second systematic. 
In the numerical work, we will combine the two errors in quadrature.

\item

The decay length (see Appendix B) of the mirror lepton is also computed. 
In Fig.~(\ref{Decaylength}), we show the contour plots of 0.15, 0.5, 1mm and 1 cm
for the decay length of $e^{M} \to l_i + \phi_{kS}$ on the ($g_{0S}, m_{M}$) plane. 
We sum over all $i$ and $k$ and set $|g_{0S}|=|g_{0S}^{\prime}|=|g_{1S}|=|g_{1S}^{\prime}|$ for simplicity. In the limit of $m_M \gg m_S$, the decay length is sensitive to neither the two scenarios mentioned above nor the neutrino mass hierarchies and the CP phases.

\begin{figure}[hptb!]
\centering
\includegraphics[width=0.5\textwidth]{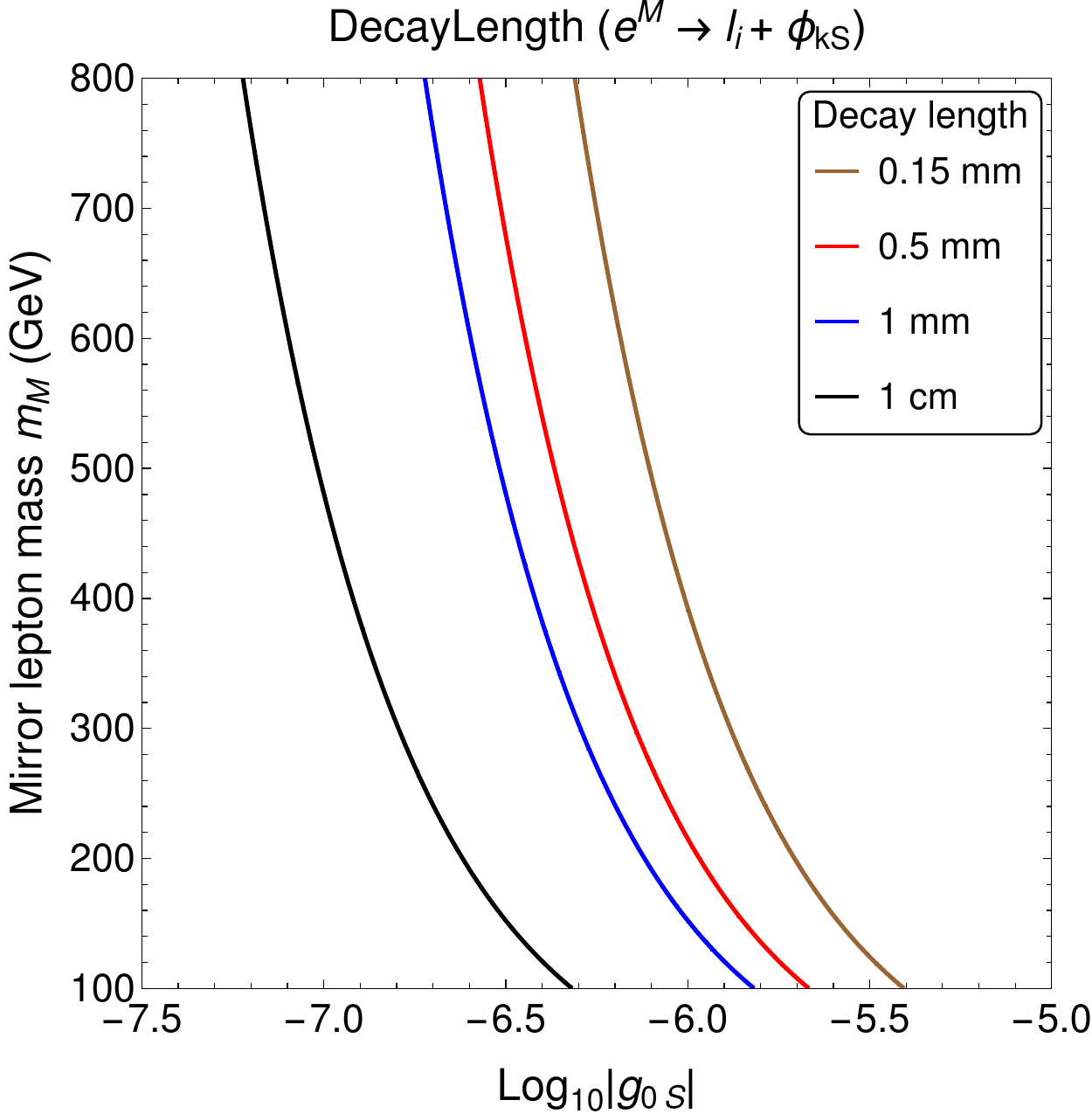} 
\caption{{\small Contour plot for decay length of $e^M \to l + \phi_{S}$ on the ($\log_{10} \vert g_{0S} \vert, m_{M}$) plane with $|g_{0S}|=|g_{0S}^{\prime}|=|g_{1S}|=|g_{1S}^{\prime}|$. }}
\label{Decaylength}
\end{figure}

\end{itemize}
%

%
\begin{figure}[hptb!]
\centering
\includegraphics[width=0.6\textwidth]{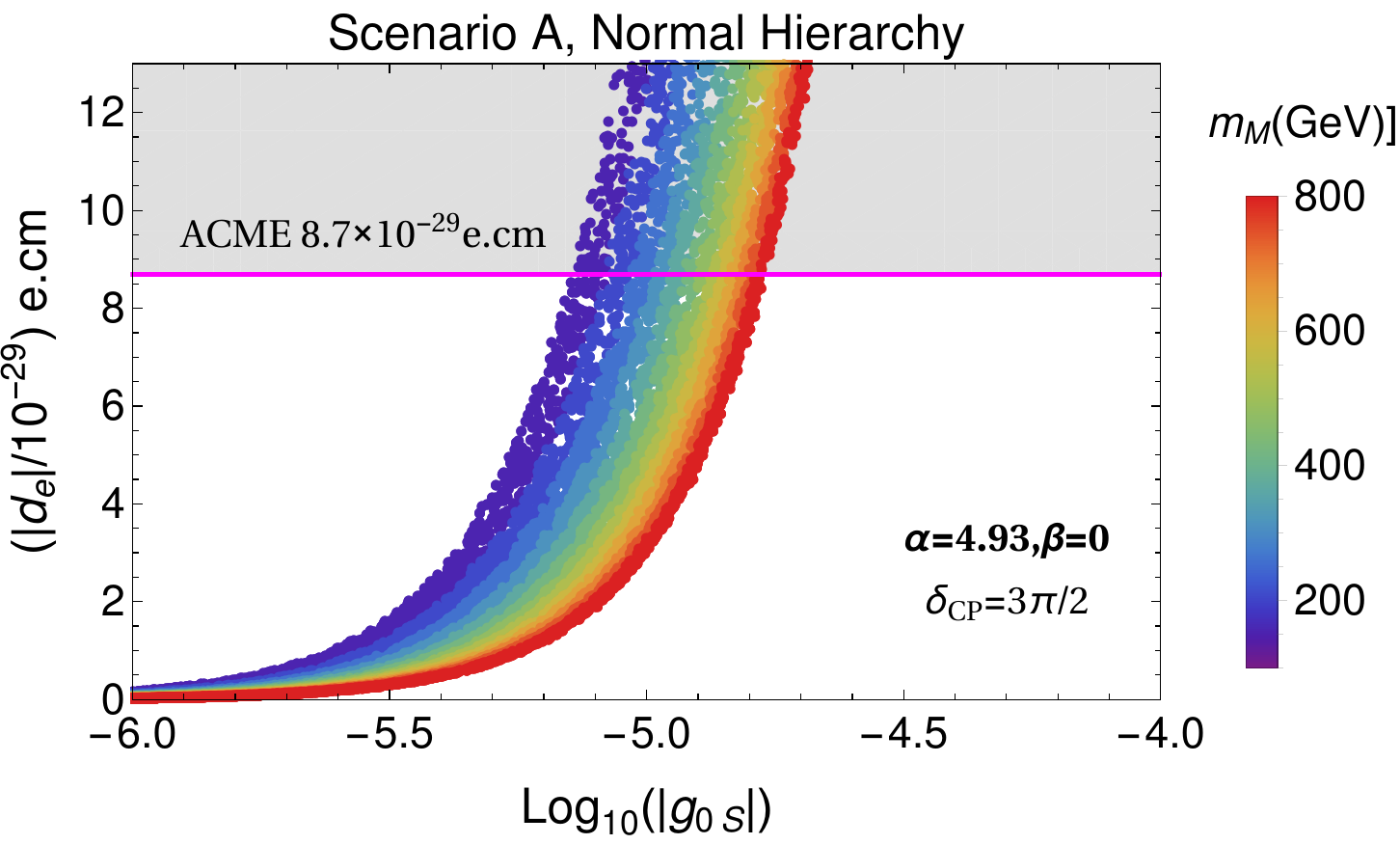} 
\caption{\small Electron EDM versus $\log_{10} \vert g_{0S} \vert$ 
in the Scenarios A and normal hierarchy for
$\alpha=4.93$, $\beta=0$ and $\delta_{\rm CP}=3 \pi /2$. 
The current upper limit of the electron EDM from the ACME Collaboration \cite{Baron:2013eja} is 
indicated by the pink line.
The color pattern represents various values of the mirror lepton mass $m_M$ in logarithmic scale. 
We set $|g_{0S}|=|g_{0S}^{\prime}|=|g_{1S}|=|g_{1S}^{\prime}|$.}
\label{eEDM-A123}
\end{figure}

In Fig.~(\ref{eEDM-A123}), we plot the electron EDM as a function of $\log_{10} \vert g_{0S} \vert$ in Scenario A and normal hierarchy 
for $\alpha=4.93$ and $\beta=0$ where $\vert J_e^A \vert$ is maximized. 
For simplicity, we set $|g_{0S}|=|g_{0S}^{\prime}|=|g_{1S}|=|g_{1S}^{\prime}|$.
Different color represents different mirror lepton mass $m_M$ as indicated by the palette at the right side of the plot. 
The pink line is the current limit of electron EDM from ACME~\cite{Baron:2013eja}. Results for inverted hierarchy are similar and will not be shown.

\begin{figure}[hptb!]
\centering
\includegraphics[width=0.6\textwidth]{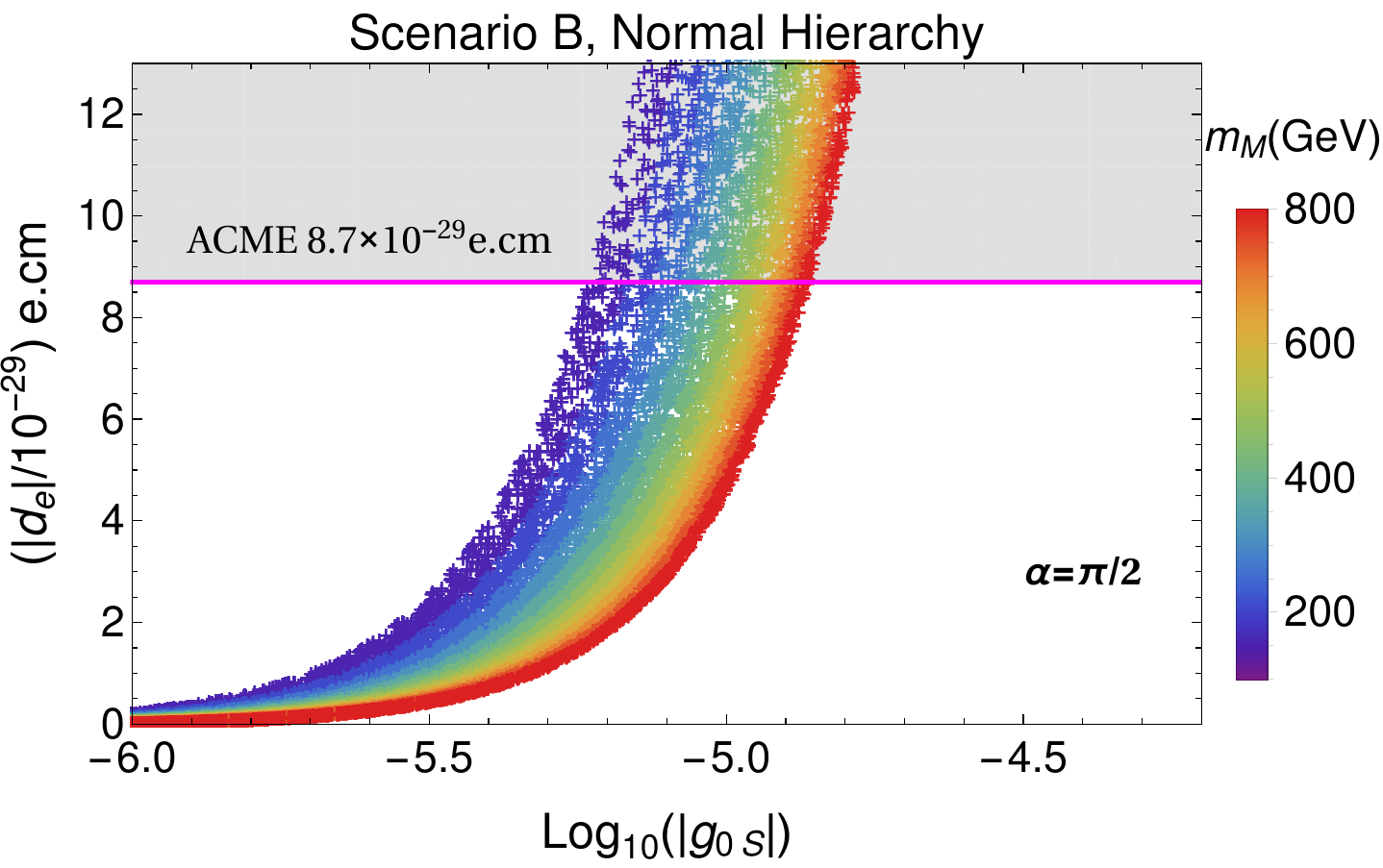} 
\caption{\small Same as Fig.~(\ref{eEDM-A123}) for Scenario B with $\alpha = \pi/2$.}
\label{eEDM-B12}
\end{figure}

Fig.~(\ref{eEDM-B12}) is the same as  Fig.~(\ref{eEDM-A123}) for Scenario B with 
$\alpha = \pi/2$ where $\vert J^B_e \vert$ is maximized.
Note that for Scenario B, from~(\ref{edmsimple}) and (\ref{JB}) in Appendix C, 
the electron EDM is independent of $\beta$ and vanishes for $\alpha=0$.

\begin{figure}[hptb!]
\centering
\includegraphics[width=0.9\textwidth]{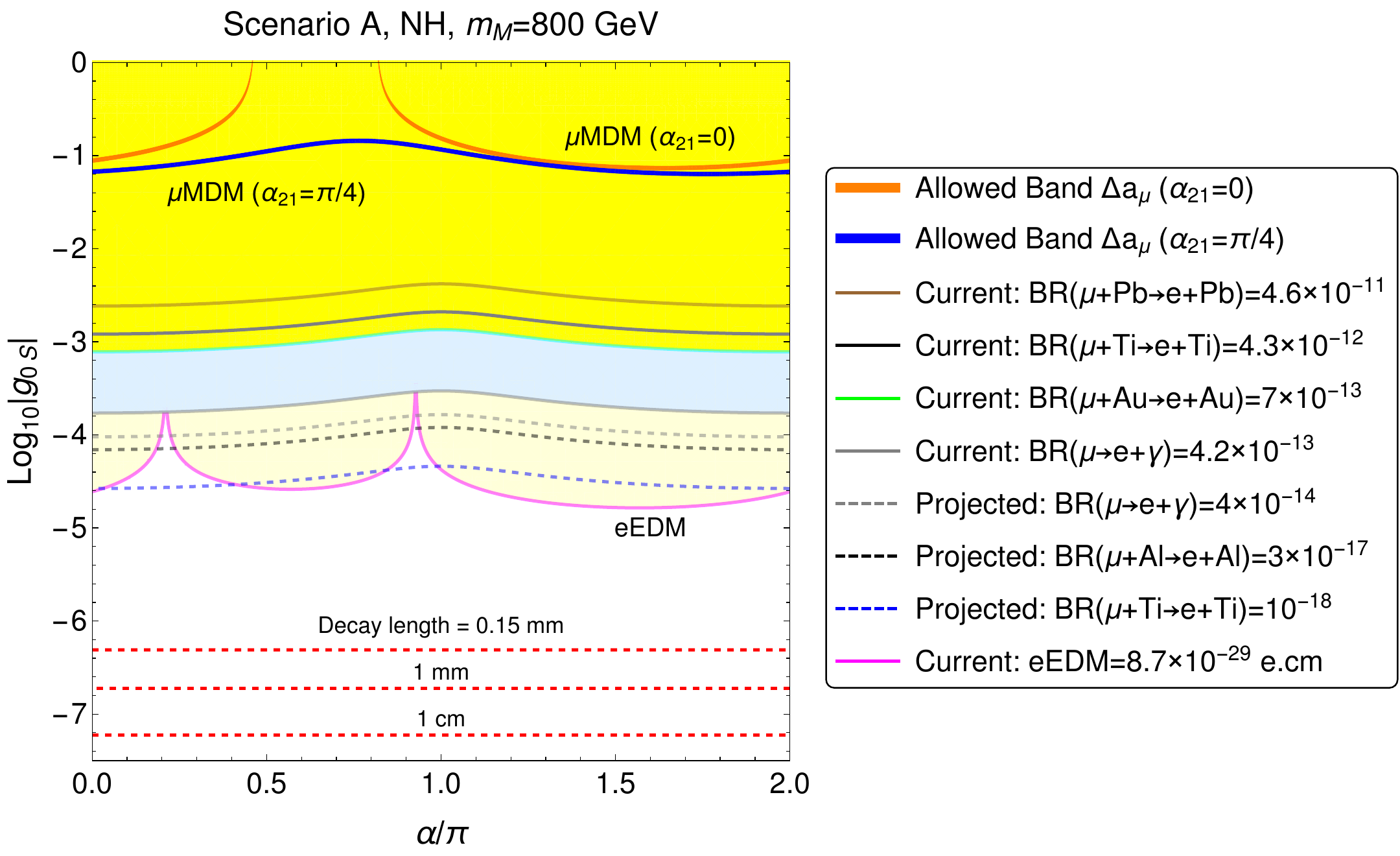} 
\caption{{\small Constraints from the current limits and project sensitivities of $\mu - e$ conversion, $\mu \to e \gamma$ and electron EDM on the magnitude of the couplings  
and their phases in normal mass hierarchy with Scenario A  for $\beta = 0$ and $\delta_{\rm CP} = 3\pi / 2$. 
The straight dashed lines are the decay length of various values for the mirror electron.
The two orange and blue bands are the allowed regions of the muon anomalous magnetic dipole moment with the Majorana phase
$\alpha_{21}=0$ and $\pi/4$ respectively. The mirror lepton mass 
$m_{M}$ is taken to be 800 GeV. }}
\label{CBA200}
\end{figure}
\begin{figure}[hptb!]
\centering
\includegraphics[width=0.8\textwidth]{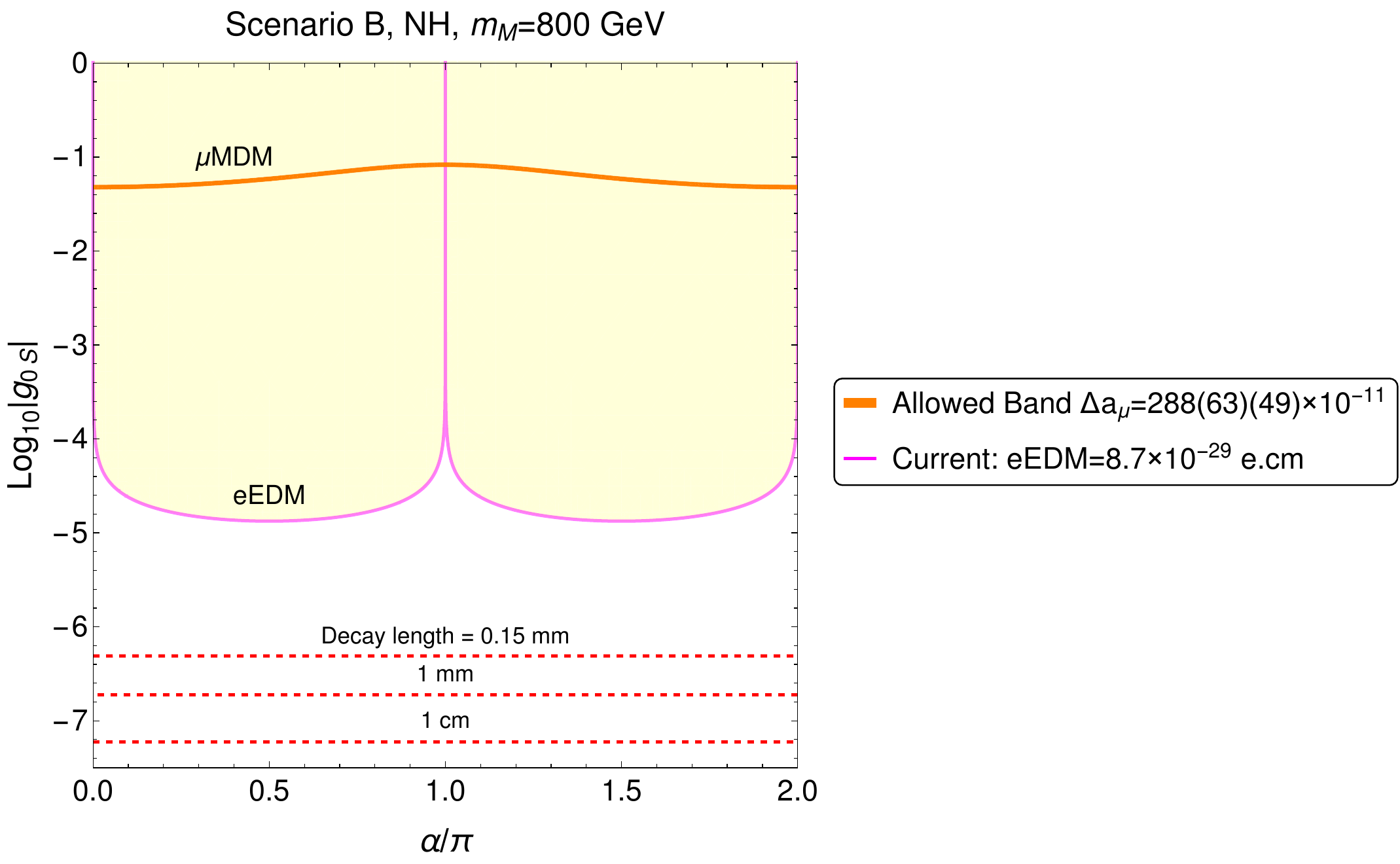} 
\caption{{\small The same as Fig.~(\ref{CBA200}) but for Scenario B.}}
\label{CBB200}
\end{figure}
In Fig.~(\ref{CBA200}) and Fig.~(\ref{CBB200}), we show the constraints on the magnitude of the couplings and the
CP phase $\alpha$ from the current  limits and projected sensitivities 
of $\mu-e$ conversion, $\mu \to e \gamma$ and electron EDM
from various experiments for normal hierarchy with $m_{M}=800$ GeV 
in both Scenario A and B respectively. We set $\beta = 0$ for simplicity.
Results for negative $\alpha$ are symmetric to those of positive $\alpha$ and will not be shown.
We note that from (\ref{muecon2muegamma}) and (\ref{muegammaScenarioB}) in Appendix E, 
both $\mu \to e \gamma$ and $\mu-e$ conversion are approximately vanishing for 
$m_M \gg m_S$ in Scenario B. 
We also note that CP phases (both Dirac and Majorana phases) can enter into the muon anomaly calculation
as in the case of MSSM case \cite{Nath:1991dn}.
The orange (blue) band in Fig.~(\ref{CBA200}) for Scenario A 
is the $3.6\sigma$ discrepancy of the muon anomaly given in (\ref{E821}) 
with the Majorana phase $\alpha_{21}=0$ ($\pi/4$).
For Scenario B, the muon anomaly does not depend on the Majorana phases.
However, they are not favored by the CLFV processes which constrains the couplings to be much smaller.

Since there is no significant difference for both neutrino mass hierarchies, 
we only show the normal hierarchy case for all plots.
Results for other values of the mirror fermion mass are qualitatively the same and will be omitted too.

\section{Conclusion}
\label{sec5}

We have studied the electron EDM in the mirror fermion model with electroweak scale non-sterile right-handed neutrinos and a horizontal $A_4$ symmetry in the lepton and scalar sectors. 
Modulo the possibility of cancellation in the various CP violation phases in different scenarios such that 
the quantity $J_e$ defined in Eq.~(\ref{Ji}) vanishes,
current experimental limit on the electron EDM imposes the most stringent constraints on the parameter space of the model, as compared with other low energy precision observables like
$\mu \to e \gamma$ and $\mu - e$ conversion in nuclei. 
However, projected sensitivities of $\mu \to e \gamma$ from MEG and of $\mu - e$ conversion experiments at Mu2e, Mu2e II, COMET and PRISM, can provide comparable if not 
more stringent constraints on the parameters of the mirror fermion model.

The region of parameter space that can ``explain" the muon anomalous magnetic dipole 
moment in the mirror fermion model is not favored by the current limits of 
these CLFV processes and the electron EDM from various experiments,
which suggest much smaller couplings of order $10^{-4}$ to $10^{-5}$. 

On the other hand, the parameter space that can be probed by current and near future experiments for these CLFV processes and the electron EDM 
is close to the region where the mirror leptons, 
when produced at the LHC, have the decay length of about 1 mm. 
Besides missing energies, the search strategies for these mirror leptons at the LHC~\cite{Chakdar:2016adj} may have to include displaced 
vertices located at distances from 1 mm to 1 cm away from the beam axises.
It is interesting to note that SM background is expected to be small in this region and signatures for mirror leptons could be distinctive. 
It is also interesting to see how, in the mirror model, low energy experiments (rare processes, electron EDM) guide the direct searches at high energy experiments (like the LHC)  
for new particles such as the mirror leptons.

For the two scenarios that we considered in this work, our results are not sensitive to the two neutrino mass
hierarchies. 

Besides EDM, the new CP violating Yukawa couplings studied here may have implications 
for leptogenesis via the asymmetry of two CP conjugate rates of mirror lepton decay
$l^M_m \to l_i \phi_{kS}$ and $\bar l^M_m \to \bar l_i \phi^*_{kS}$. 
Unfortunately, the small magnitudes of these new Yukawa couplings deduced from this work 
indicate that the asymmetry generated might be too small. 
This issue is interesting and deserved for further investigation.

To complete the story, one might want to extend the $A_4$ symmetry to the quark sector. 
The experimental constraints from the well established quark mixings in the CKM model 
must then be faced \cite{quarkA4}.  Needless to say, it is also of interest to study the neutron 
EDM in the mirror fermion model. Work on this is now in progress and will be reported elsewhere.

\vfill

\section*{Acknowledgments}
This work is supported by the Ministry of Science and Technology (MoST) of Taiwan under
grant 104-2112-M-001-001-MY3. 


\section*{Appendix}

In this Appendix, we collect some useful formulas used in this work.

\subsection*{(A) $C^{ij}_L$ and $C^{ij}_R$}

These coefficients were computed in \cite{Hung:2015hra} and we collect their expressions here 
for convenience.
\bea
\label{CL}
C^{ij}_L & = & + \frac{e}{16 \pi^2} \sum_{k=0}^3 \sum_{m=1}^3 
\left\{
\frac{1}{m^2_{l^M_m}}
\left[
m_i \mathcal{U}^{R \, k}_{jm} \left( \mathcal{U}^{R \, k}_{im} \right)^* 
+ m_j \mathcal{U}^{L\, k}_{jm} \left( \mathcal{U}^{L \, k}_{im} \right)^* 
\right]
{\cal I} \left( \frac{m^2_{\phi_{kS}}}{m^2_{{l^M_m}}} \right) \right. \nonumber \\ 
&\;& \;\;\;\;\;\;\;\;\;\; \;\;\;\;\;\;\; \;\;\;\;\;\;\;+ \left. \frac{1}{m_{l^M_m}} \mathcal{U}^{R \, k} _{jm} \left( \mathcal{U}^{L\, k}_{im} \right)^* {\cal J} \left( \frac{m^2_{\phi_{kS}}}{m^2_{{l^M_m}}} \right)
\right\} \;\; ,\\
\label{CR}
C^{ij}_R & = & + \frac{e}{16 \pi^2} \sum_{k=0}^3 \sum_{m=1}^3 
\left\{
\frac{1}{m^2_{l^M_m}}
\left[
m_i \mathcal{U}^{L\, k}_{jm} \left( \mathcal{U}^{L\, k}_{im} \right)^* 
+ m_j \mathcal{U}^{R \, k}_{jm} \left( \mathcal{U}^{R \, k}_{im} \right)^* 
\right]
{\cal I} \left( \frac{m^2_{\phi_{kS}}}{m^2_{{l^M_m}}} \right) \right. \nonumber \\ 
&\;& \;\;\;\;\;\;\;\;\;\; \;\;\;\;\;\;\; \;\;\;\;\;\;\;+ \left. \frac{1}{m_{l^M_m}} \mathcal{U}^{L\, k}_{jm} 
\left( \mathcal{U}^{R\, k}_{im} \right)^* {\cal J} \left( \frac{m^2_{\phi_{kS}}}{m^2_{{l^M_m}}} \right)
\right\} \; \; ,
\eea
where $m_{i,j}$, $m_{l^M_m}$ and $m_{\phi_{kS}}$ are the SM leptons, mirror leptons and scalar singlets masses respectively, with the subscripts $i,j,m$ being the generation indices. 
In the calculation~\cite{Hung:2015hra}, we have assumed $m_{l^M_m} \gg m_{i,j}$ and 
set $m_{i,j} \to 0$ in the loop functions 
${\cal I}(r)$ and ${\cal J}(r)$, which are simply given by
\bea
\label{I}
{\cal I}(r) & = & \frac{1}{12 (1 - r)^4} \left[ - 6 r^2 \log r + r ( 2 r^2 + 3 r - 6 ) + 1 \right] \;\; , \\
\label{J}
{\cal J}(r) & = & \frac{1}{2 (1 - r)^3} \left[ - 2 r^2 \log r + r ( 3 r - 4) + 1 \right] \; \; .
\eea
Note that ${\cal I}(0) = 1/12$ and ${\cal J}(0) = 1/2$.

\subsection*{(B) Decay Length of Mirror Leptons}

For an unstable relativistic particle, its decay length $l$ is given by $l = \beta \gamma c \tau$, where $\beta=v/c$ is its velocity, $\gamma = 1/(1 - \beta^2)^{1/2}$ its dilation factor, and $\tau = (\sum \Gamma)^{-1}$ its total lifetime with $\Gamma$ being its partial width.

The decay rate for $l^M_m \to l_i + \phi_{kS}$ is given by 
\bea
\label{decaylength}
\Gamma (m \to ik) & = & \frac{1}{32 \pi}m_{l^M_m} 
\left( 1 - \left( \frac{m_{l_i} + m_{\phi_k}}{m_{l^M_m}} \right)^2 \right)^{1/2}
\left( 1 - \left( \frac{m_{l_i} - m_{\phi_k}}{m_{l^M_m}} \right)^2 \right)^{1/2} \nonumber\\
& \times & 
\left\{ 
\left( 1 + \frac{m_{l_i}^2 - m_{\phi_k}^2}{m_{l^M_m}^2} \right) 
\left( ( {\cal U}^{L \, k }_{im} )^* {\cal U}^{L \, k}_{im} + ( {\cal U}^{R \, k }_{im} )^* {\cal U}^{R \, k}_{im} \right) 
\right. \nonumber \\
&& + \left. \left( 2 \frac{m_{l_i}}{m_{l^M_m}} \right) 
 \left( ( {\cal U}^{L \, k }_{im} )^* {\cal U}^{R \, k}_{im} + ( {\cal U}^{R \, k }_{im} )^* {\cal U}^{L \, k}_{im} \right) 
\right\} \; .
\eea

\subsection*{(C) Muon Anomaly}

For Scenario A, the muon anomaly is given by

\bea
\Delta a_{\mu}^{A} &\approx &\frac{1}{16 \pi^2}
\left\{\frac{m_{\mu}^2}{6 m^{2}_{M}} \Big[ |g_{0S}|^{2}+|g^{\prime}_{0S}|^{2} + 2\,(|g_{1S}|^{2}+|g^{\prime}_{1S}|^{2})\Big] \right. \nonumber \\
 &&  + \left. \frac{m_{\mu}}{4\, m_{M}}\,\Big[ |g_{0S}|\,|g^{\prime}_{0S}| 
 \Big(C_{3}\cos(\frac{\alpha_{21}}{2}-\alpha) +C_{4}\sin(\frac{\alpha_{21}}{2}-\alpha) \right. \nonumber
 \\ 
 && \left. \hspace{1cm} + C_{5}\sin(\frac{\alpha_{21}}{2}+\delta_{\rm CP}-\alpha)+C_{6}\cos(\frac{\alpha_{21}}{2}+\delta_{\rm CP}-\alpha)\Big) \right. \nonumber \\
 && +\left. 2 \, |g_{1S}|\,|g^{\prime}_{1S}| \, \cos(\beta) \, \Big(C_{3}\cos(\frac{\alpha_{21}}{2}) + C_{4}\sin(\frac{\alpha_{21}}{2}) \right. \nonumber \\
 && \left. \hspace{1cm} +C_{5}\sin(\frac{\alpha_{21}}{2}+\delta_{\rm CP})+C_{6}\cos(\frac{\alpha_{21}}{2}+\delta_{\rm CP})\Big)\Big] 
\right \}  \; ,
\eea
where 
\bea
C_{3}&=& \frac{1}{\sqrt{3}}\,\big(-c_{12} c_{23}+c_{12}s_{23} +2\,s_{12}c_{13}\big)\,,\nonumber \\
C_{4}&=& c_{12} \big(s_{23}+c_{23}\big)\,,
\nonumber \\
C_{5}&=& s_{12} s_{13}\big(c_{23}-s_{23}\big)\,,\nonumber \\
C_{6}&=& \frac{1}{\sqrt{3}}\,s_{12} s_{13}\big(c_{23}+s_{23}\big)\,.
\eea

For Scenario B, we have
\bea
\Delta a_{\mu}^{B} &\approx &\frac{1}{16 \pi^2}
\left\{\frac{m_{\mu}^{2}}{6m^{2}_{M}} \Big[|g_{0S}|^{2}+|g^{\prime}_{0S}|^{2} + 2\,(|g_{1S}|^{2}+|g^{\prime}_{1S}|^{2})\Big] \right. \nonumber \\
 && \quad \quad \; + \left. \frac{m_{\mu}}{2m_{M}}\,\Big[ |g_{0S}||g_{0S}^{\prime}| \cos(\alpha)\,
 +2\,|g_{1S}||g_{1S}^{\prime}| \,\cos(\beta) \Big] 
\right \}  \; . 
\eea
Note that the muon anomaly depends on the Majorana phase $\alpha_{21}$ for Scenario A but not for Scenario B 
in the mirror fermion model we are studying.

Exact analytical formula of the anomalous MDM for lepton $l_i$ can be found in \cite{Hung:2015hra}. 

%
\begin{table}[b!]
\begin{tabular}{lcc}
\hline
Nuclei \, \, & $\Gamma_{\rm capt}$ ($10^{6}$ s$^{-1}$) & $D$ \\
\hline\hline
$^{48}_{22}$Ti &  2.59 & 0.0864\\
$^{197}_{79}$Au & 13.07 & 0.189\\
$^{208}_{82}$Pb & 13.45  & 0.161 \\
\hline
\end{tabular}
\caption{SM values of the capture rates 
(in unit of $10^6 \, {\rm s}^{-1}$~\cite{muoncapturerate}) 
and the dimensionless overlap integrals (evaluated under the assumption that the proton and neutron distributions within each nuclei are the same~\cite{Kitano:2002mt}) for titanium, gold and  lead.
 }
\label{tabconvrate}
\end{table}

\subsection*{(D) $\mu - e$ Conversion and Radiative Decay $\mu \to e \gamma$ }

The branching ratios for the $\mu - e$ conversion rate and $\mu \to e \gamma$ are related as~\cite{Hung:2017voe}
\be
{\rm Br}( {\mu N \to e N} ) = \frac{\Gamma^{\gamma^*}_{\rm conv}}{\Gamma_{\rm capt}} \approx \pi D^2 \frac{\Gamma_\mu}{\Gamma_{\rm capt}} 
{\rm Br}( \mu \to e \gamma)
\label{muecon2muegamma}
\ee
where $\Gamma_{\rm capt}$ and 
$D$ are the capture rate and overlap integral of the nuclei $N$ respectively. Their values for different nuclei are listed in Table~\ref{tabconvrate} for convenience. $\Gamma_\mu$ is the total width of the muon. 
Detailed analytical expressions for $\mu - e$ conversion in nuclei can be found in \cite{Hung:2017voe}.

For Scenario A, we have
\bea
\Gamma^A \left( \mu \to e \gamma \right) &\approx &\frac{1}{16 \pi} m^3_{\mu} \left( 1 - \frac{m^2_{e}}{m^2_{\mu}} \right)^3 \,\,\left(\frac{e}{32\pi^{2}\,m_{M}}\right)^{2}\Big(C_{7}+C_{8}\cos(\delta_{\rm CP})+C_{9}\sin(\delta_{\rm CP})\Big) \nonumber \\ 
 && \hspace{15pt}\times \Big[|g_{0S}|^{2}\,|g^{\prime}_{0S}|^{2}+2\,|g_{1S}|^{2}\,|g^{\prime}_{1S}|^{2}\big(1+\cos(2\beta)\big) \nonumber \\
&&\;\;\;\;\;\;\;\;\;\;\;\quad\quad \quad\quad\quad\quad\quad+\,4\,|g_{0S}||g_{0S}^{\prime}||g_{1S}||g_{1S}^{\prime}|\cos(\alpha)\cos(\beta)\Big] \; ,
\label{muegammaScenarioA}
\eea
where 
\bea
C_{7}&=& \frac{1}{3} \Big(2-c_{12}^{2}c_{23}s_{23}\big(s_{13}^{2}+2\big)+3\,c_{12}c_{13}s_{12}(c_{23}-s_{23})+s_{12}^{2}c_{23}s_{23}(2\,s_{13}^{2}+1)\Big) \; ,
\nonumber \\
C_{8}&=&\frac{1}{3}\, \Big(s_{13}(c_{23}+s_{23})\big(c_{12}^{2}c_{13}+3\,c_{12}s_{12}(s_{23}-c_{23})-2\,c_{13}s_{12}^{2}\big)\Big) \; , \nonumber \\
C_{9}&=&\frac{1}{\sqrt{3}}\,c_{12}s_{13}\Big(c_{12}c_{13}(c_{23}-s_{23})+s_{12}\Big) \; .
\eea
We note that the CP violation phases enter into the decay rate of $\mu \to e \gamma$ in Eq.~(\ref{muegammaScenarioA}).

For Scenario B, we have 
\be
\Gamma^{B}\left( \mu \to e \gamma \right) \approx 0 \; .
\label{muegammaScenarioB}
\ee
Perhaps this null result needs some explanations.  
Note that the amplitude for $l _{i} \to l_{j} +\gamma$ ($i \neq j$) is proportional to the off-diagonal elements 
of $C^{ij}_{L,R}$. However, under the set up of the parameters space discussed in Sect.~\ref{sec4}, 
$C^{ij}_{L}$, for example, has the following form
\bea
C^{ij}_L  &\approx &  \sum_{k=0}^3 \sum_{m=1}^3 \Big\{ a \,\mathcal{U}^{R \, k}_{jm} \Big(\mathcal{U}^{R \, k}_{im} \Big)^*  + b\, \mathcal{U}^{L\, k}_{jm} \Big( \mathcal{U}^{L \, k}_{im} \Big)^* + c\, \mathcal{U}^{R \, k} _{jm} \Big( \mathcal{U}^{L\, k}_{im} \Big)^* \Big\} 
\eea 
where $a, b, c$ are some constants related to the masses. 
In Scenario B, one has 
\bea
C^{B,ij}_L  
&=& \sum_{k=0}^3 \Big\{ U^{\dagger}_{\rm PMNS} \Big[ a\, M^{\prime k} \Big(M^{\prime k}\Big)^{\dagger} + b\, M^{ k} \Big(M^{k}\Big)^{\dagger} + c\, M^{\prime k} \Big(M^{ k}\Big)^{\dagger} \Big] U_{\rm PMNS} \Big\}_{ji} 
\eea 
One can easily check $\sum_{k=0}^3 M^{\prime k} \Big(M^{\prime k}\Big)^{\dagger}$, $\sum_{k=0}^3 M^{ k} \Big(M^{k}\Big)^{\dagger}$ and $\sum_{k=0}^3 M^{\prime k} \Big(M^{ k}\Big)^{\dagger}$ are diagonal in $i,j$.
Since $U_{\rm PMNS}$ is unitary,  this implies $C^{B,ij}_L$ is diagonal. Similarly, $C^{B,ij}_R$ is also diagonal.
Thus $\Gamma^{B}\left( \mu \to e \gamma \right) \approx 0$, and there is no $\mu-e$ conversion as well in Scenario B, according 
to Eq.~(\ref{muecon2muegamma}).

In general, the charged lepton flavor violating processes can depend on Majorana phases~\cite{Gavela:2009cd}.
However, in both Scenario A and B of the mirror fermion model that we are studying, one can check that these Majorana phases 
do not enter in $\mu \to e \gamma$.

Exact formulas for the rate of $l_i \to l_j + \gamma$ can be found in  \cite{Hung:2015hra}.


\end{document}